\documentclass[twocolumn]{aastex631}

\usepackage{soul}
\usepackage{amsmath}
\usepackage{lipsum} 
\usepackage{multirow}
\usepackage{xspace, fontawesome}

\PassOptionsToPackage{hyphens}{url}\usepackage{hyperref}

\newcommand{\Mpch}{\,h^{-1}\text{Mpc}}

\newcommand{\githubmaster}{\href{https://github.com/victoriaono/variational-diffusion-cdm}{\faGithub}\xspace}

\begin{document}

\title{Debiasing with Diffusion: Probabilistic reconstruction of Dark Matter fields from galaxies with CAMELS}

\author{Victoria Ono}
\affiliation{Harvard-Smithsonian Center for Astrophysics, 60 Garden Street, Cambridge, MA 02138, USA}
\affiliation{Harvard University, 17 Oxford Street, Cambridge, MA 02138, USA}
\email{victoriaono@college.harvard.edu}

\author[0000-0002-9542-2913]{Core Francisco Park}
\affiliation{Harvard-Smithsonian Center for Astrophysics, 60 Garden Street, Cambridge, MA 02138, USA}
\affiliation{Department of Physics, Harvard University, 17 Oxford Street, Cambridge, MA 02138, USA}

\author[0000-0001-5139-612X]{Nayantara Mudur}
\affiliation{Harvard-Smithsonian Center for Astrophysics, 60 Garden Street, Cambridge, MA 02138, USA}
\affiliation{Department of Physics, Harvard University, 17 Oxford Street, Cambridge, MA 02138, USA}

\author[0000-0001-7899-7195]{Yueying Ni}
\affiliation{Harvard-Smithsonian Center for Astrophysics, 60 Garden Street, Cambridge, MA 02138, USA}

\author[0000-0002-6069-2999]{Carolina Cuesta-Lazaro}
\affiliation{Harvard-Smithsonian Center for Astrophysics, 60 Garden Street, Cambridge, MA 02138, USA}
\affiliation{The NSF AI Institute for Artificial Intelligence and Fundamental Interactions Massachusetts Institute of Technology, Cambridge, MA 02139, USA}
\affiliation{Department of Physics, Massachusetts Institute of Technology, Cambridge, MA 02139, USA}

\author[0000-0002-4816-0455]{Francisco Villaescusa-Navarro}
\affiliation{Center for Computational Astrophysics, Flatiron Institute, 162 5th Avenue, 
New York, NY 10010, USA}
\affiliation{Department of Astrophysical Sciences, Princeton University, 4 Ivy Lane, Princeton, 
NJ 08544 USA}

\begin{abstract}

Galaxies are biased tracers of the underlying cosmic web, which is dominated by dark matter components that cannot be directly observed. 
Galaxy formation simulations can be used to study the relationship between dark matter density fields and galaxy distributions. 
However, this relationship can be sensitive to assumptions in cosmology and astrophysical processes embedded in the galaxy formation models, that remain uncertain in many aspects. 
In this work, we develop a diffusion generative model to reconstruct dark matter fields from galaxies. The diffusion model is trained on the CAMELS simulation suite that contains thousands of state-of-the-art galaxy formation simulations with varying cosmological parameters and sub-grid astrophysics. 
We demonstrate that the diffusion model can predict the unbiased posterior distribution of the underlying dark matter fields from the given stellar mass fields, while being able to marginalize over uncertainties in cosmological and astrophysical models. Interestingly, the model generalizes to simulation volumes $\approx 500$ times larger than those it was trained on, and across different galaxy formation models. Code for reproducing these results can be found at \url{https://github.com/victoriaono/variational-diffusion-cdm} \githubmaster. 

\end{abstract}


\section{Introduction} \label{sec:intro}

Within the standard $\Lambda$CDM paradigm, hierarchical structure formation arises from the gravitational growth and collapse of the initial density inhomogeneities and forms the large-scale cosmic web of today's universe. 
85\% of the matter content is composed of dark matter (DM), whose nature remains one of the most enigmatic questions in astrophysics due to the absence of direct observations. 
Large observational surveys such as DESI~\citep{DESI2016arXiv161100036D}, Euclid~\citep{Laureijs2011arXiv1110.3193L}, Roman~\citep{Spergel2015arXiv150303757S}, and
Rubin~\citep{LSST2009arXiv0912.0201L} are devoted to mapping the cosmos by observing millions of galaxies at different wavelengths which will serve as biased tracers of the underlying dark matter density fields with the goal of improving our understanding of the nature and constituents of the Universe. 



The cosmic web is sensitive to the laws and constituents of the Universe. However, the dominant component of the cosmic web, dark matter, is not directly observable. Therefore, one needs to infer the distribution of the cosmic web solely based on biased observable tracers. In this paper, we present a probabilistic approach to reconstruct dark matter density fields from such observations. 
The reconstructed cosmic web can be used for two purposes: 1) to study field-level cosmology and extract information from voids and filaments that may not be detectable from the galaxy distribution alone, and 2) to identify regions with low ratios of stellar to dark matter mass, as those are the regions where we expect dark matter signatures to be larger, and can aid in constraining the nature of dark matter.

On large scales, the clustering of galaxies can be described by a perturbative bias expansion of the dark matter density \citep{Desjacques_2018}, where the complexity of galaxy formation physics is contained in a small set of expansion coefficients referred to as biased parameters. Alternatively, one can describe galaxy biasing in the context of the Effective Field Theory of Large Scale Structures \citep{Senatore_2015}, which provides a systematic framework for modeling galaxy clustering based only on symmetries and scale separation. Currently, these models can only accurately reproduce the galaxy power spectrum on scales larger than $10 \Mpch$ \citep{Ivanov_2021}.

On smaller scales, perturbation theory breaks down, and the clustering of galaxies is affected by nonlinear structure formation and astrophysical processes such as supernova or AGN feedback, requiring hydrodynamical simulations for theoretical predictions in this regime.
Over the past decade, galaxy formation simulations have made significant progress (see \cite{vogelsberger20} for a recent review) to more accurately study the relationship between the observed galaxy distributions and the underlying dark matter fields.

However, due to limited resolution, cosmological simulations usually adopt coarse-grained sub-grid models to effectively describe the small-scale astrophysical processes such as star formation, supernova feedback, black hole evolution and AGN feedback, and there are still large theoretical uncertainties lying in those sub-grid models that lead to different predictions made by different galaxy formation simulations. 

To encapsulate those uncertainties, the Cosmology and Astrophysics with MachinE Learning Simulations (CAMELS) project \citep{villaescusa2021camels,Ni2023arXiv230402096N} generated more than 8000 state-of-the-art galaxy formation simulations that widely explore the variations in cosmological and astrophysical parameters within different galaxy formation models, to provide a broad training set and testing ground for machine-learning algorithms designed for cosmological studies.

Machine learning (ML) is a powerful tool to learn complex high dimensional distributions, such as the mapping between observable fields (e.g. galaxies, neutral hydrogen gas) and the underlying cosmic web from simulations.
For example, previous work has tried to use convolutional neural network models to infer the dark matter field from the galaxy density field \citep{hong2021revealing} and from 21cm maps \citep{Villanueva-Domingo2021ApJ...907...44V}.  
However, these models are deterministic and cannot generate the posterior distribution of different samples of the cosmic web given the observable fields.
It is important to develop probabilistic generative models, such as \cite{mudur2022denoising} and \cite{park2023probabilistic}, which can encapsulate our uncertainties in how galaxies connect to the underlying dark matter distribution. 
These models enable us to address questions such as the likelihood of a certain dark matter halo mass being associated with a given galaxy distribution.

In this work, we develop a diffusion generative model trained on the CAMELS simulation suites to reconstruct the underlying DM fields from stellar density fields.
The primary goal of the diffusion model is to capture the relationship between the stellar fields and DM fields and predict the unbiased posterior distribution of the DM fields conditioned on the given stellar field, $p(x_\mathrm{DM}|x_\mathrm{stars})$.
By training on the CAMELS suite, the diffusion model can account for the uncertainties inherent in the astrophysical processes assumed by galaxy formation models, as well as the specific cosmological parameters utilized in the simulations. 
We also train the diffusion model using various training sets from CAMELS and evaluate its robustness across different galaxy formation models. 
Additionally, we apply the trained diffusion model to large simulations of IllustrisTNG-300 to showcase its extrapolation performance on out-of-distribution data.

The paper is structured as follows. 
Section~\ref{sec:Methodology} presents the diffusion model along with the dataset from CAMELS used for training and testing purposes. 
Section~\ref{sec:Results} conducts a series of validations to assess the performance and robustness of the diffusion model. 
Finally, Section~\ref{sec:discussion} provides a summary of the paper.

\begin{figure*}
    \centering
    \includegraphics[width=1.0\textwidth]{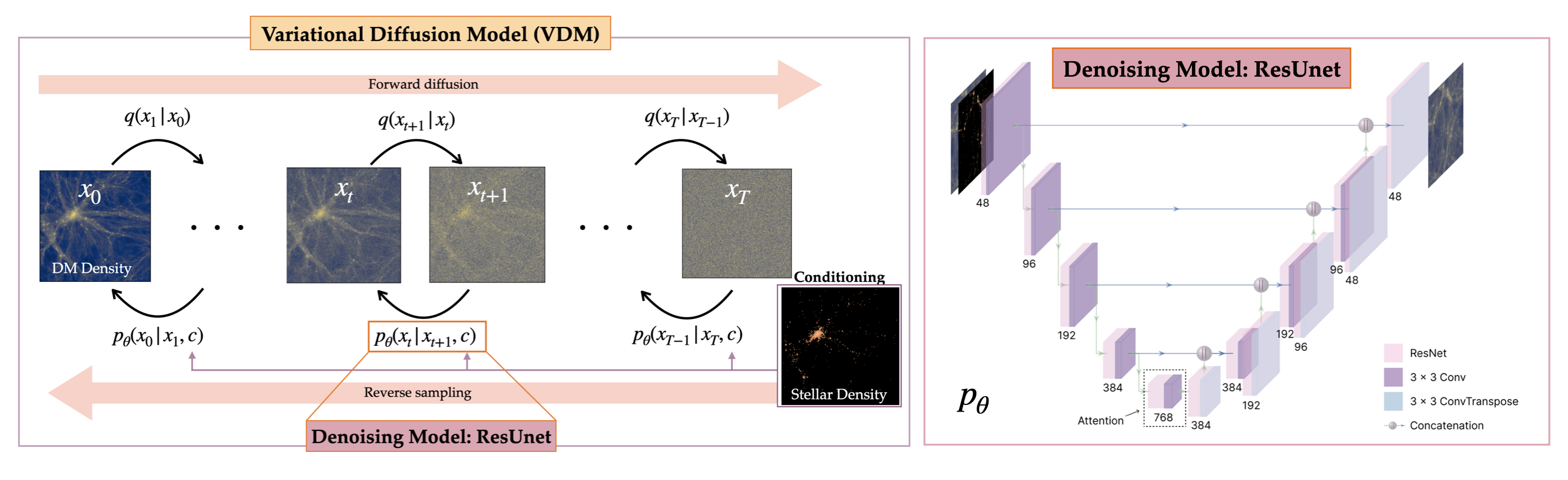}
    \caption{Schematic overview of the conditional diffusion model used to model the posterior distribution of dark matter density fields given the stellar density field. The left panel illustrates the diffusion process, and the right panel shows the details of the convolutional-neural-network-based denoising model.}
    \label{fig:schematic}
\end{figure*}

\section{Methodology}
\label{sec:Methodology}

In this section, we describe the dataset used, the model architecture, and training methods. 

\begin{table*}
\centering
\begin{tabular}{| c | c | c | c | c | c | c| c| c |  }
    \hline
    Simulation suites & Set & Number of Simulations & $\Omega_m$ & $\sigma_8$ & $A_{\rm SN1}$ & $A_{\rm SN2}$ & $A_{\rm AGN1}$ & $A_{\rm AGN2}$ \\
    \hline
    \hline
    \multirow{4}{*}{ASTRID / IllustrisTNG / SIMBA} 
    & LH & 1000 & $0.1\mbox{-}0.5$ & $0.6\mbox{-}1.$ & $0.25\mbox{-}4.$ & $0.5\mbox{-}2.$ & $0.25\mbox{-}4.$ & $0.5\mbox{-}2.$ \\
    & 1P & 61 & $0.1\mbox{-}0.5$ & $0.6\mbox{-}1.$ & $0.25\mbox{-}4.$ & $0.5\mbox{-}2.$ & $0.25\mbox{-}4.$ & $0.5\mbox{-}2.$  \\ 
    & CV & 27 & $0.3$ & $0.8$ & $1$ & $1$  & $1$ & $1$\\
    \hline
\end{tabular}
\caption{Summary of the simulation sets, which share the same design between the suites ASTRID, IllustrisTNG, and SIMBA suites. $A_{\rm SN1}$, $A_{\rm SN2}$, $A_{\rm AGN1}$, $A_{\rm AGN2}$ represent the value of subgrid physics parameters controlling stellar and AGN feedback. The LH set varies parameters in a latin-hypercube, whereas the 1P set has variations of each parameter at a time, with all others fixed to their fiducial values. The CV set has all parameters fixed to the fiducial values, but varies the random seed of the initial conditions.}
\label{tab:table2}
\end{table*}

\subsection{Dataset}
In this work, we use 2D maps from the CAMELS Multifield Dataset \citep{villaescusa2022camels} to model the connection between stellar mass and dark matter densities produced by three different suites of hydrodynamical simulations: ASTRID, IllustrisTNG, and SIMBA.
Table~\ref{tab:table2} provides a summary of the simulations in each suite.

CAMELS simulations have a box volume of $(25 h^{-1}{\rm Mpc})^3$. From each simulation at $z=0$, CAMELS Multifield Dataset produces 15 paired maps representing the stellar and dark matter surface density in a region with dimensions $25\times25\times5~(h^{-1}{\rm Mpc})^3$ that is projected along the third axis. 
We keep the image size to the original $256 \times 256$ pixels, corresponding to $25 \Mpch$ on both sides. 
The resolution of our maps is therefore of $\approx 0.1 \Mpch$.

We use the Latin Hypercube (LH) set of each of the three simulation suites to train our models. 
Each LH sets contains $1000$ independent simulations spanning a wide range of cosmological and astrophysical parameters, reflecting the uncertainties of cosmology and the complex astrophysical processes taking place in our current understanding of galaxy formation. 
We augment the training set with random flips and permutations of the input and output images. 

The parameters varied in the LH set for each simulation suite are $\Omega_{\text{m}}, \sigma_8$ (cosmological), $A_{\text{SN1}}, A_{\text{AGN1}}, A_{\text{SN2}}$ and $A_{\text{AGN2}}$ (astrophysical), and their ranges are: $0.1 \leq \Omega_m \leq 0.5, 0.6 \leq \sigma_8 \leq 1.0, 0.25 \leq (A_{\rm SN1}, A_{\rm AGN1}) \leq 4.00$, and $0.5 \leq (A_{\rm SN2}, A_{\rm AGN2}) \leq 2.0$. 
$A_{\text{SN}}$ and $A_{\text{AGN}}$ control the strength of supernova and AGN feedback respectively, with their exact physical meaning differing across different galaxy formation models (see \cite{Ni2023arXiv230402096N} for detailed descriptions).

\begin{figure*}[ht]
    \centering
    \includegraphics[width=0.9\textwidth]{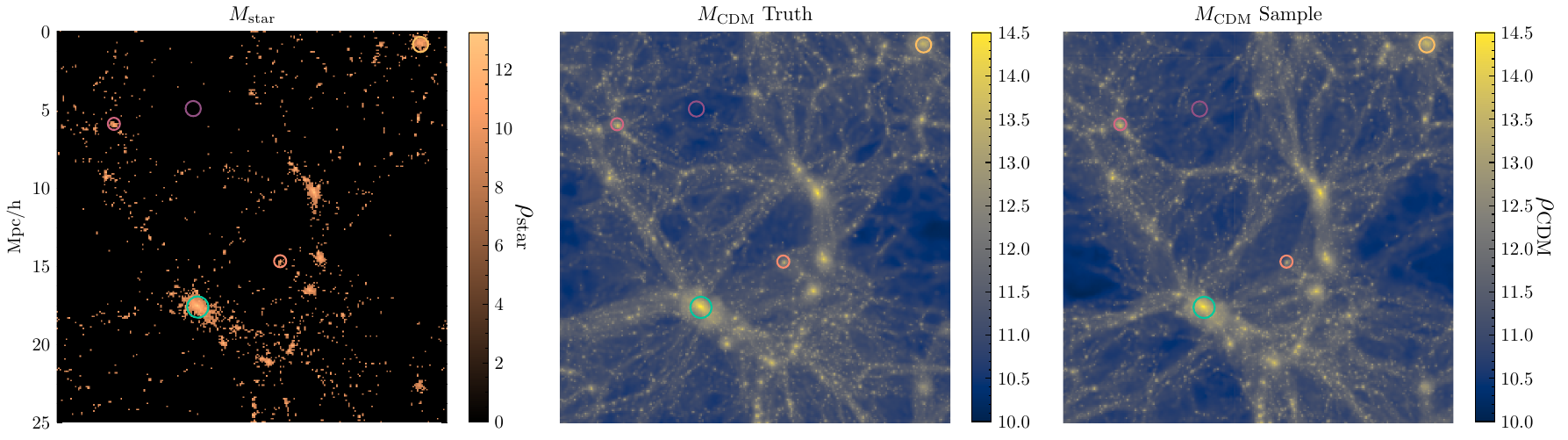}
    ~
    \includegraphics[width=0.9\textwidth]{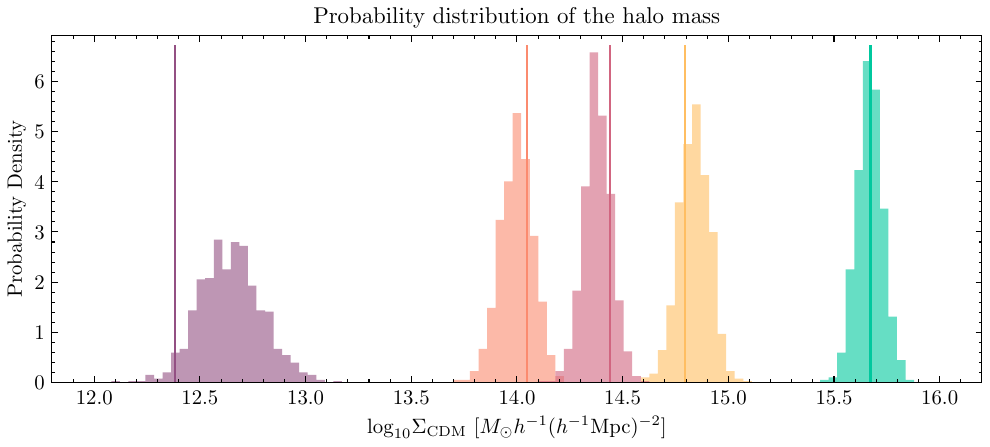}
        \caption{\textbf{Top row}: Input stellar field, corresponding true dark matter density field, and a sample dark matter density field from the ASTRID-trained model, with circles drawn for selected regions ranging from a void to a massive star cluster. \textbf{Bottom row}: Probability distribution of the halo mass (defined to be the sum of the pixels in the region) from 1000 samples corresponding to each selected region, with the solid line representing the true mass. The model is able to predict closer to the true mean as the total mass in the region increases, and its decrease in variance demonstrates the model's increase in confidence.
        }
    \label{fig:mass-regions}
\end{figure*}



We use the different simulation suites to assess the generalization capabilities of the trained diffusion models.  In particular, the 1P set varies one parameter (from the fiducial value in the CV set) at a time for each simulation. 
Testing on this dataset would clearly indicate how capable the model is at marginalizing over each astrophysical parameter. 
On the other hand, the CV set contains 27 simulations with the same fiducial values of cosmological and astrophysical parameters, but varied random seeds of the initial conditions that are designed to quantify the level of cosmic variance on different cosmological and astrophysical probes. The fiducial values are $\Omega_m=0.3, \sigma_8=0.8, A_{\rm SN1}= A_{\rm AGN1}=A_{\rm SN2}=A_{\rm AGN2}=1$.


\subsection{Diffusion Model}

Figure~\ref{fig:schematic} provides a schematic overview of the diffusion model used in this study. 
We use the variational diffusion model developed by \cite{kingma2021variational} with a denoising architecture similar to the U-Net \citep{ronneberger2015u} to model the posterior of DM density fields given stellar fields. The conditional diffusion model predicts the conditional probability of the dark matter field $x_\mathrm{DM}$ from the stellar field $x_\mathrm{stars}$,

\begin{equation}
    p(x_\mathrm{DM}|x_\mathrm{stars}) = \frac{p(x_\mathrm{DM})p(x_\mathrm{stars}|x_\mathrm{DM})}{p(x_\mathrm{stars})}.
\end{equation}

Our diffusion model generates a target DM density field in $T=250$ refinement steps. It begins with a random Gaussian noise field $x_\mathrm{DM}^T \sim \mathcal{N}(0,1)$, and iteratively denoises it according to a learned conditional probability $p_\theta(x_\mathrm{DM}^{t-1}|x_\mathrm{DM}^{t}, x_\mathrm{stars})$ to ultimately generate a sample $x_\mathrm{DM}^0 \sim p(x_\mathrm{DM}|x_\mathrm{stars})$. During training, the denoising model takes as input $\{x_\mathrm{DM}^{t},x_\mathrm{stars}, t\}$ and estimates the noise that was added to the image at that timestep. 

In the forward diffusion process, we progressively add noise to an image by sampling from 
$q(x_\mathrm{DM}^{t} \vert x_\mathrm{DM}^{0}) = \mathcal{N}(\alpha_t x_\mathrm{DM}^{0}, \sigma_t^2\mathbf{I})$,
where $\alpha_t$ and $\sigma^2_t$ are functions of $\gamma_t$, the variance schedule, which we assume to be a linear function of time and whose free parameters are learned during training. Noise is added to the sample according to this schedule and in a variance-preserving way, i.e. $\alpha_t^2 = \text{sigmoid}(-\gamma(t))$. 

The loss function we optimize is the variational lower bound of the marginal likelihood. The objective thus is to minimize a bound to the posterior $p_\theta(x_\mathrm{DM}| x_\mathrm{stars})$.

A detailed ablation study of this model is presented in Appendix~\ref{app:ablation}.

\bigskip

\subsection{Training}
For our denoising model, we use a hierarchical U-Net \citep{ronneberger2015u}-like architecture with 4 blocks of double convolution followed by strided downsampling layers. We employ group normalization \citep{wu2018group} and residual connections \citep{kaiming_resnet} in each block, and use the AdamW optimizer \citep{adamW} with a learning rate of $1\times 10^{-4}$ and the CosineAnnealingWarmRestarts learning rate scheduler \citep{loshchilov2017sgdr}. We also initialize the learned linear noise schedule with $\gamma(t)=26.6\:t-13.3$. We train the model using the PyTorch Lightning framework \citep{Falcon_PyTorch_Lightning_2019} with a batch size of $12$ for $60000$ gradient steps, using the LH set as our training data. 

We select the model with the lowest mean-squared error (MSE) in the validation set, calculated as 
\begin{equation*}
\mathrm{MSE} = \mathbb{E}_{x^\mathrm{Sample}_\mathrm{DM} \sim
 p(x_\mathrm{DM}|x_\mathrm{stars})} \left(x^\mathrm{Sample}_\mathrm{DM} - x^\mathrm{True}_\mathrm{DM} \right)^2.
\end{equation*}

\section{Results}\label{sec:Results}
In this section, we conduct an exhaustive set of tests to demonstrate the trained model's performance and robustness:

\begin{enumerate}
    \item In~\ref{subsection:CV-set}, we showcase the model's performance when the cosmological and astrophysical parameters are set to their fiducial values, using the CV set for testing the model.
    \item In~\ref{subsection:1P-set}, we demonstrate that the model generalizes to different cosmological and astrophysical parameters by varying one of them at a time, using the 1P set.
    \item In~\ref{subsection:cross-test}, we show that the model generalizes across galaxy formation simulations after training the diffusion model on a single simulation suite and testing on the others.
    \item In~\ref{subsection:increase_volume}, we test the model's ability to recover the large-scale dark matter distribution by deploying a model trained on small volumes to a simulation with an $\approx 500$ times larger volume.
\end{enumerate}


\subsection{Predictions for the CV set}
\label{subsection:CV-set}
\begin{figure*}[ht]
    \centering
    \includegraphics[width=\textwidth]{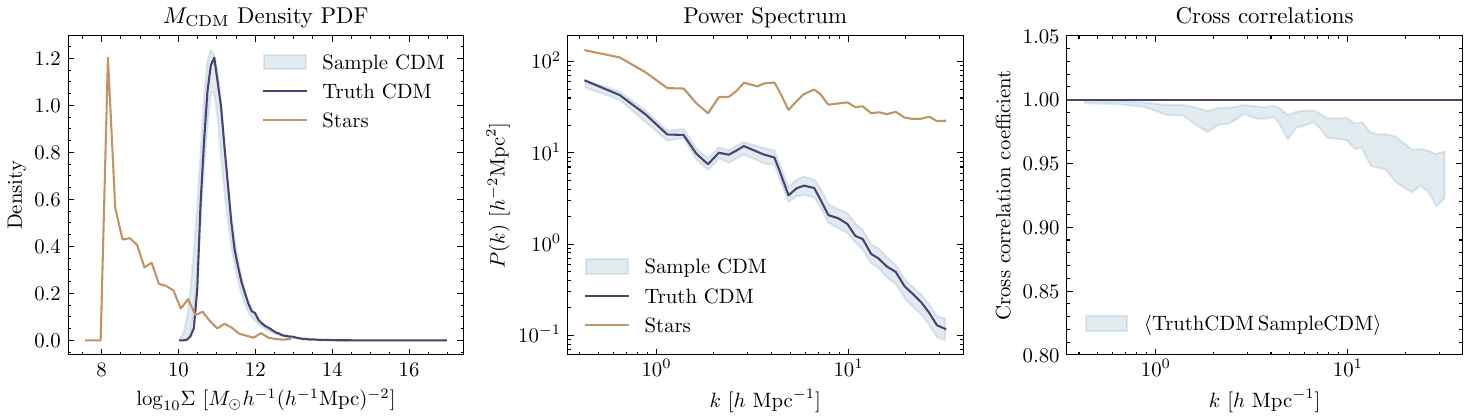}
    \caption{Summary statistics of the single stellar field and its corresponding DM density field from the CV set. \textbf{Left panel:} density histogram of stars (copper), true dark matter (solid blue line) and dark matter inferred by the diffusion model (light blue). Note that we show the 10-90th percentiles of $100$ samples from the posterior distribution. \textbf{Middle panel:} Power spectra for star, true dark matter and sampled dark matter fields. \textbf{Right panel:} Cross correlations between true and sampled dark matter fields. All panels show good agreement between the summary statistics of the true and sampled dark matter fields, demonstrating the model's ability to well reproduce the statistical properties of the cosmic web.}
    \label{fig:cv-summaries}
\end{figure*}


In this section, we test the diffusion model trained on maps of the LH set of the ASTRID simulation suite on maps from ASTRID's CV set, whose cosmological and astrophysical parameter values are fixed to the fiducial ones. 

The top panel of Figure~\ref{fig:mass-regions} illustrates one posterior sample of the DM density field in comparison to the true DM density field. The model can qualitatively reproduce the expected features of the dark matter cosmic web of nodes, filaments and voids. 

The bottom panel in Figure~\ref{fig:mass-regions} demonstrates how the posterior samples can be used to estimate posteriors of quantities of interests, in particular the mass of selected regions in the density field. 
Here, the mass is calculated as the sum of pixel values in the selected circular region. The regions for which posteriors are shown are highlighted in the maps above, ranging from a void to a massive star cluster. As expected, the posteriors are broader in regions with low density of stars. 

We also train the same diffusion model on the CV set, instead of the LH set, to test how much uncertainty is introduced in the posterior samples by implicitly marginalizing over the varying astrophysical and cosmological parameters in the LH set.
We do not see a significant reduction in variance in the model trained on the CV set, given the large range of variations in the cosmological and astrophysical priors used to generate the LH set simulations. This reduction in variance would also come at the cost of poorer generalization capabilities. We therefore train all our models on the LH set.

Figure~\ref{fig:cv-summaries} shows a quantitative comparison between the summary statistics of the true DM fields and those of the generated ones from the corresponding stellar fields. The results are shown for one sample of the CV set.
From left to right, we show the statistics of density histograms, the 2D power spectra and the 2D cross power spectra, with shaded regions obtained from the posterior distribution of generated DM map 100 samples, quantifying the 10th to 90th percentile uncertainties of the model predictions.  

The left panel of Figure~\ref{fig:cv-summaries} shows good agreement between the 1D histogram of the true DM field and the sampled DM fields.
The middle panel of Figure~\ref{fig:cv-summaries} compares the power spectra of the true DM map and generated samples. We see that the diffusion model can reproduce the expected dependence of the power spectrum with scale, and that the true power falls within the uncertainty of the samples. 


Finally, the right panel of Figure~\ref{fig:cv-summaries} compares the cross-correlation coefficient between the true and sampled DM fields. 
Cross-correlation coefficients are calculated by taking the cross-power spectra between the sampled DM and true DM field and dividing by their auto-power spectra:
\[
R_{\langle \text{True} \, \text{Sample} \rangle}(k) = \frac{P_{\langle \text{True} \, \text{Sample} \rangle}(k)}{\sqrt{P_{\text{True}}(k)}\sqrt{P_{\text{Sample}}(k)}}
\]
$R(k)$ measures the correlation between the phases of the modes of the two fields. 
In an ideal scenario where the true DM and sampled DM fields are perfectly correlated, the cross correlation between true DM and sample DM would be $1$ across all the scales. 
We find that the cross-correlations between the sampled and true DM fields are always higher than $0.9$, demonstrating that the model is able to well reproduce the statistical properties of the cosmic web.

In Appendix~\ref{app:posterior_uncertainties}, we show how the posterior variance is lowest (highest) at pixels with high (low) non-zero stellar mass.

\begin{figure*}[ht]
    \centering
    \includegraphics[width=\textwidth]{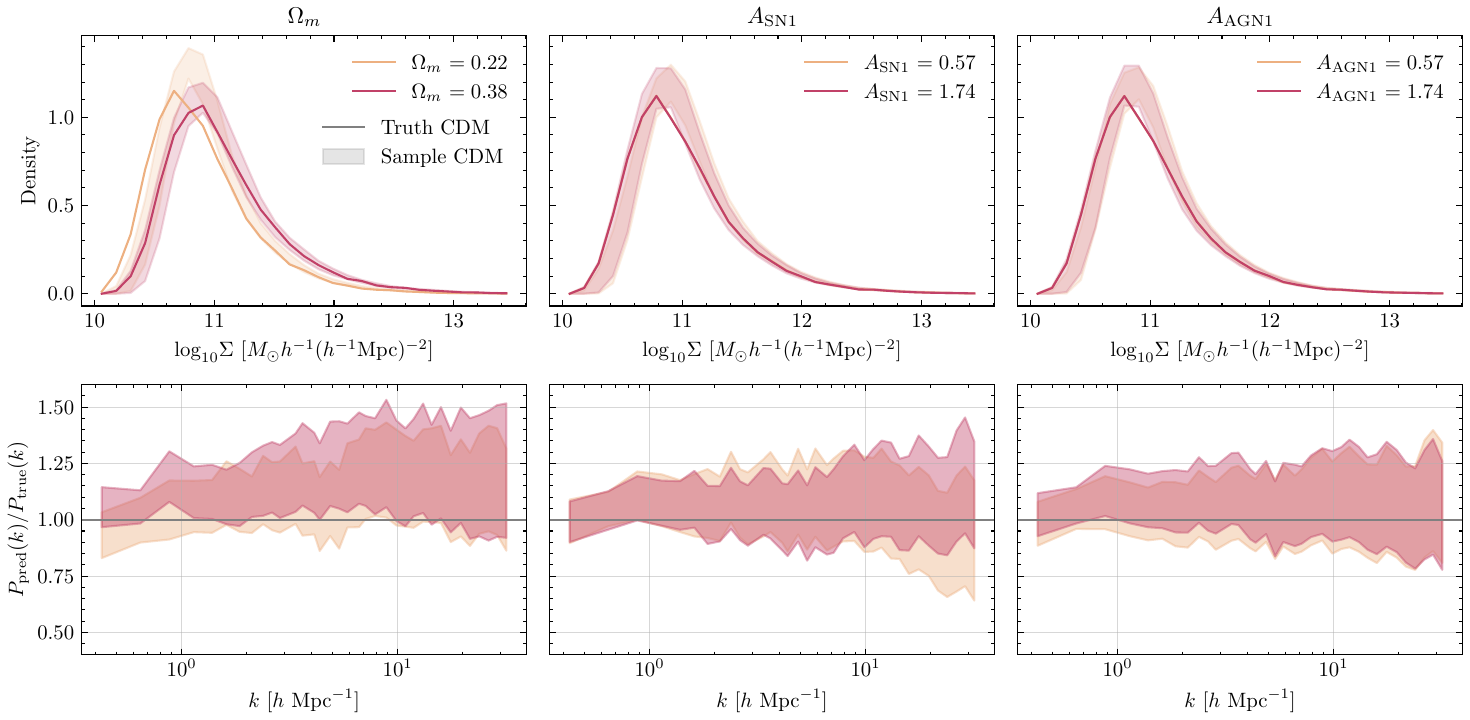}
    \caption{Summary statistics of the density fields when varying $\Omega_m$, $A_\mathrm{SN1}$ and $A_\mathrm{AGN1}$, while fixing all other parameters to their fiducial values. 
    For each parameter variation, we use one single stellar map as the input and generate 100 DM samples from that map.
    \textbf{Top row}: Comparison of the density field PDFs. \textbf{Bottom row}: Power spectrum ratio between the sampled fields and the true DM field. The solid lines in the top panel give the true DM distributions from simulations, while the shaded regions correspond to 10-90th percentiles based on 100 generated DM samples from the single input stellar field. Note that the dark matter projected density distribution does not vary noticeably with the astrophysical parameters $A_\mathrm{SN1}$ and $A_\mathrm{AGN1}$, and the two solid lines overlap. The model is able to capture the overall trends well for all three parameters, though its uncertainty increases at smaller scales.}
    \label{fig:1p-summaries}
\end{figure*}

\begin{figure*}[ht]
    \centering
    \includegraphics[width=0.32\textwidth]{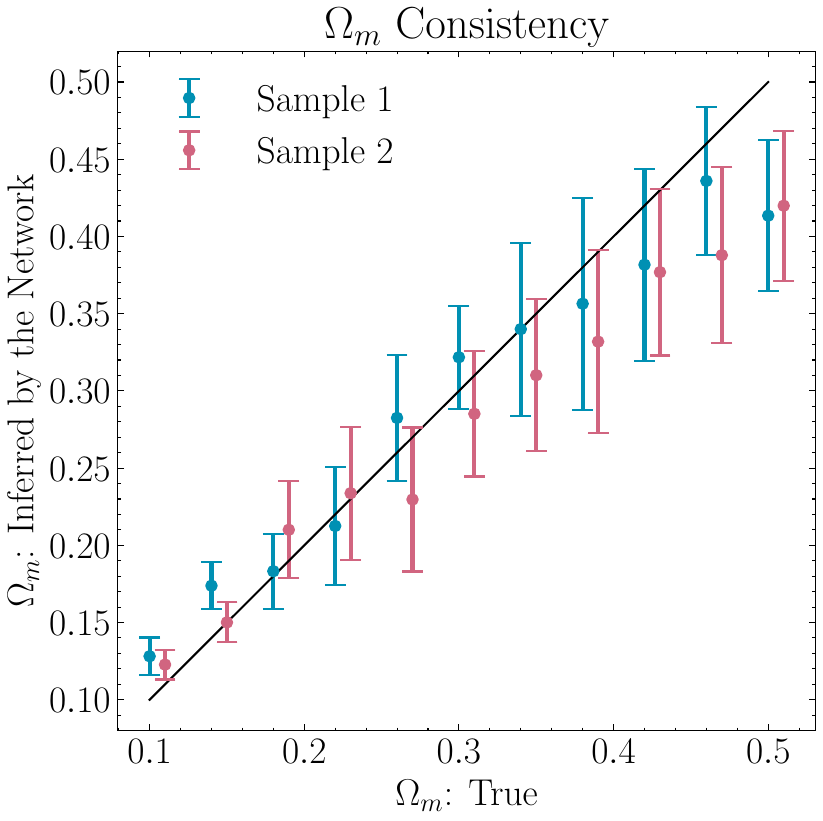}
    \includegraphics[width=0.32\textwidth]{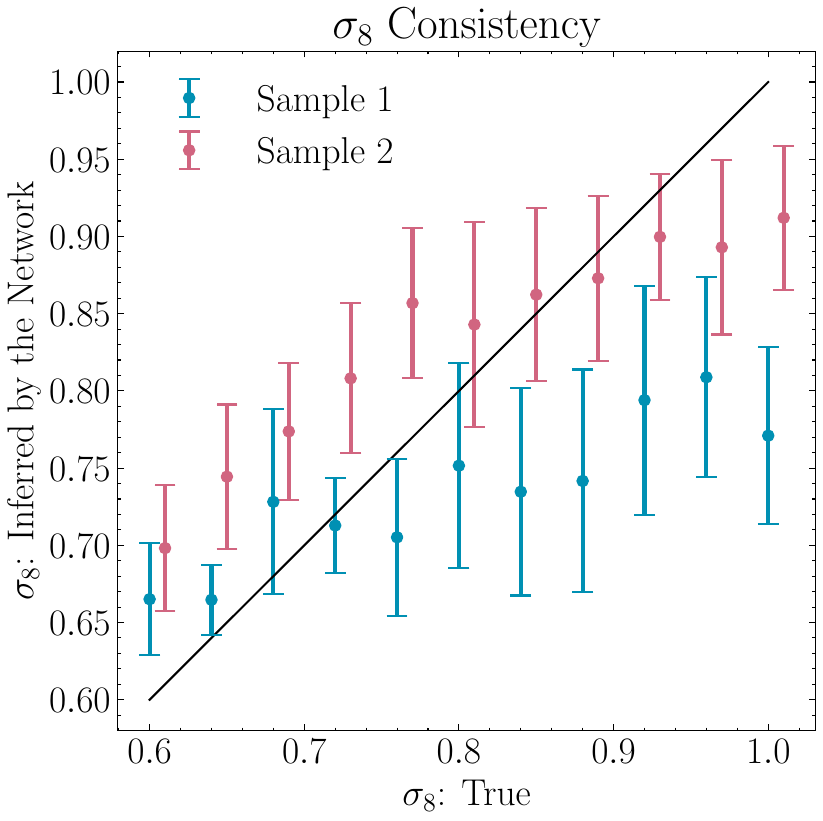}
    \caption{Predictions of the parameter inference network in \cite{2021arXiv210909747V} for 10 samples from the diffusion model $p_\theta(x_\mathrm{DM}|x_\mathrm{stars})$ for 2 independent stellar fields, denoted by the blue and red colors, for each of the 22 1P parameters with varying $\Omega_m$ and $\sigma_8$. We add a slight offset on the x-axis to help distinguish the two input fields. Each point and its error bars denote the mean and the standard deviation of the NN's parameter predictions over all 10 posterior samples of the input field corresponding to the true parameter value on the x axis. The values of $\Omega_m$ inferred by the neural network have a strong correlation with the true parameters, despite the fact that the diffusion model is not conditioned on cosmology. }
    \label{fig:1p-infnet}
\end{figure*}

\subsection{Varying cosmological and astrophysical parameters}
\label{subsection:1P-set}

Figure~\ref{fig:1p-summaries} shows a quantitative comparison between the generated DM density fields and the true ones based on the simulations in the 1P set that separately vary the parameters $\Omega_m$, $A_{\rm SN1}$ and $A_{\rm AGN1}$. 
For each selected parameter, we randomly choose a single stellar map from that simulation as the input to generate 100 DM samples.

As shown in the first column of the upper panel, the DM distribution is more sensitive to the cosmological parameter $\Omega_m$.
We show that in both low and high $\Omega_m$ scenarios, the diffusion model can effectively capture the dependence on $\Omega_m$ and reproduce the trends in the DM density distribution accordingly.

In the lower panel of Figure~\ref{fig:1p-summaries}, we show the ratio of the predicted power spectra to the true one. We find that the model tends to overpredict the power across all scales for large variations in $\Omega_m$ for this particular sample, although this is not a general trend, as can be seen in Figure~\ref{fig:cv-summaries}. 
 

For varying $A_{\rm SN1}$ and $A_{\rm AGN1}$ scenarios, the stellar density maps will look very different on small scales due to the varying strength in supernova and AGN feedback, whilst the underlying true dark matter density field will be largely unaffected. 
We see that the model is able to reproduce this behavior and generalizes well for both low and high parameter values, exhibiting its potential to marginalize over baryonic effects. We do, however, see more uncertainty at smaller scales, consistent with our findings from other summary statistics.

The consistency of the density PDFs and power spectra with varied cosmological and astrophysical parameters demonstrates that the trained diffusion model can well capture the clustering properties of the underlying DM fields based on the stellar fields, while marginalizing over the cosmological and astrophysical parameters. 

Finally, we assess the cosmological information contained in the reconstructed dark matter density fields. In particular, we use the parameter inference networks presented in
\cite{2021arXiv210909747V} to predict the cosmological parameters $\Omega_m$ and $\sigma_8$ for the reconstructed dark matter density fields. The network is trained to return a mean prediction as well as a standard deviation that indicates the network's uncertainty on its prediction.

For this test, we select two input stellar fields for each parameter in the 1P set with $\Omega_m$ and $\sigma_8$ varying, and generate 10 samples for each field from $p_\theta(x_\mathrm{DM}|x_\mathrm{stars})$. We then pass the generated dark matter density fields through the parameter inference network and examine the consistency between the parameters inferred by the neural network and the true cosmological parameters in Figure \ref{fig:1p-infnet}. 

The network-inferred values of $\Omega_m$ are close to the truth and have an average error of $14.6\%$, while it struggles with the $\sigma_8$ prediction. Since the parameter inference networks were trained on a subset of IllustrisTNG DM LH fields while we are testing our model on the ASTRID 1P set, we also find a slight offset in the inferred value of $\sigma_8$ when we test the networks on the true ASTRID 1P fields (see Figure \ref{fig:infet-true} in the Appendix). 

In Figure 2 of \cite{2021arXiv210909747V}, the parameter inference network trained on input stellar fields was able to predict $\Omega_m$ with an average error of $19.8\%$, in line with our findings.
\subsection{Generalizing beyond the training set} 
\label{subsection:cross-test}
To test the model's ability to generalize over different galaxy formation simulations, we train three independent models, using maps from the IllustrisTNG, SIMBA and ASTRID LH sets, and test each model using maps of the CV set from each of the simulation suites. 
In Figure~\ref{fig:imshows}, we show the generated samples from the same index of each CV set, which corresponds to having the same initial seed and parameter values.\footnote{Note that the definitions of the astrophysical parameters are different for each simulation suite.}

\begin{figure*}
    \centering
    \includegraphics[width=1.0\textwidth]{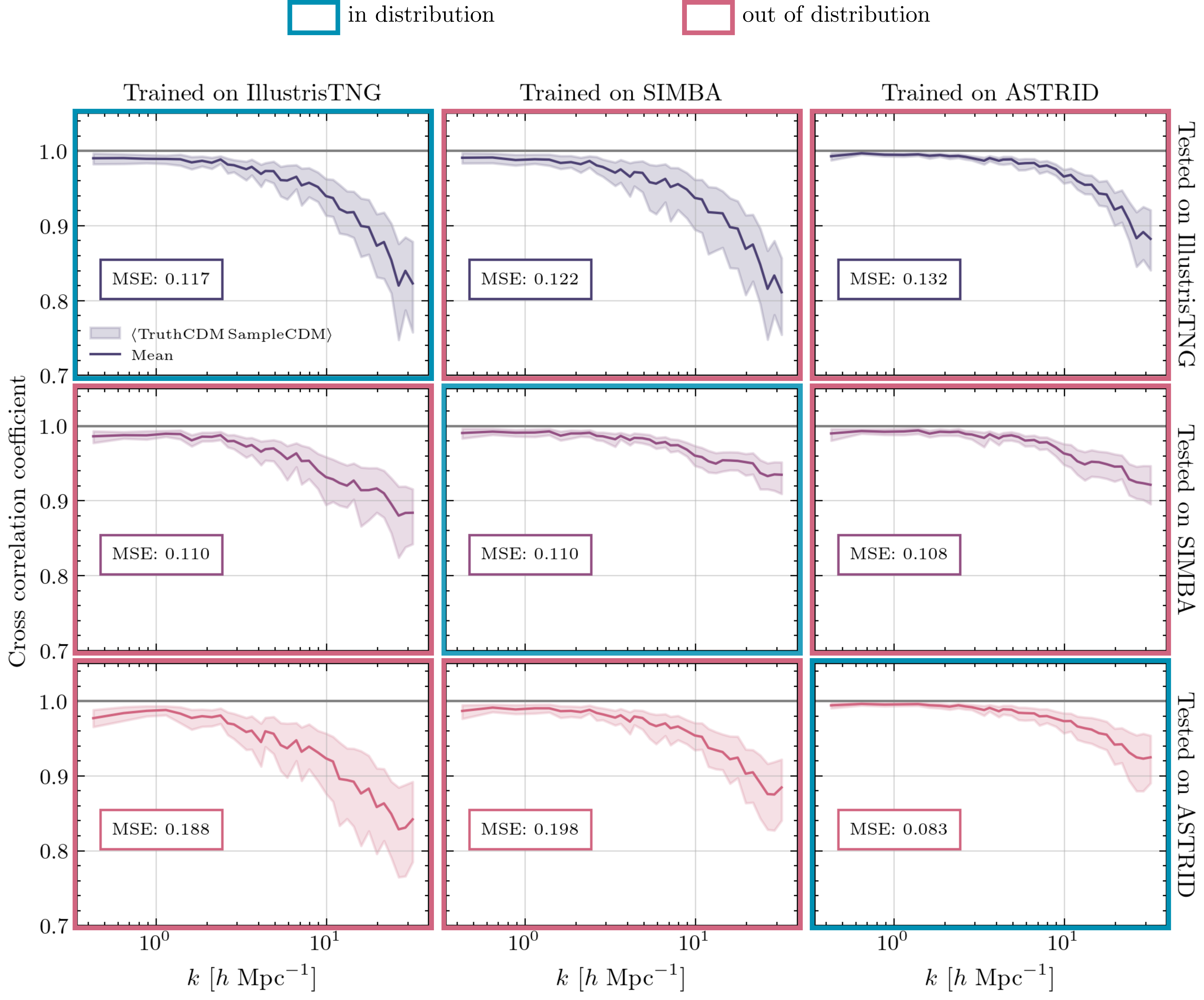}
    \caption{Cross correlations between true and 100 sampled DM fields given the same conditioning across three models trained and tested on each dataset, together with the average MSE. The shaded region in each panel represents the 10-90th percentiles of the cross-correlation coefficients for all samples, and the solid line represents their mean coefficient. We observe that the ASTRID-trained model is able to reproduce this quantity the best among all suites, while IllustrisTNG-trained model produces the lowest coefficients and largest variances. Overall, all different combinations produce cross-correlation coefficients higher than $0.8$, demonstrating that each model can generalize well across galaxy formation models.}
    \label{fig:cross-matrix}
\end{figure*}

We also quantitatively compare the cross-correlation coefficients between the true DM fields and those of the generated ones for each of these models in Figure~\ref{fig:cross-matrix}. 
As a further comparison metric, we use 100 DM samples generated from the diffusion model and calculate the averaged MSE of the true DM field and the generated samples. 



As shown in Figure~\ref{fig:cross-matrix}, 
all three diffusion models perform well when tested on simulations conducted by the same galaxy formation model (the blue boxes marked as in-distribution), and they also demonstrate robust generalization when tested on simulations carried out by other galaxy formation models (the red boxes marked as out-of-distribution). 
The generated DM fields consistently reproduce the pattern of the true cosmic web across all the scales, maintaining cross-correlation coefficients consistently higher than 0.8 for any pair.

The effective reproduction of DM density fields beyond each model's training set is facilitated by the large overlap of data distribution between the simulation suites, as illustrated in Figure~\ref{fig:LH-CV}, which shows the ratio of stellar mass to dark matter density power spectra for each LH set. 
Despite the distinct galaxy formation models employed by the three simulation suites, resulting in different mappings between the star and DM fields, the overall similarities suggest why each model can effectively generalize to every other galaxy formation model.

Comparing the models trained on the three different simulation suites, the model trained on the ASTRID suite performs the best with the minimum MSE, while the model trained on IllustrisTNG yields the largest averaged MSE. 
We speculate that the good performance of the ASTRID-trained model is due to the higher mass resolution of the ASTRID simulation for star particles. 
In the galaxy formation model employed by ASTRID, each star-forming gas particle can be split and spawn four star particles, as compared to a single star particle in IllustrisTNG and SIMBA.
Consequently, the stellar density maps within the ASTRID suite exhibit a less sparse distribution compared to the other two simulation suites, as depicted in the first column of Figure~\ref{fig:imshows}. 
The effective increased spatial resolution in ASTRID results in stellar density fields that have tighter correlations with the underlying dark matter fields, explaining the better performance of the ASTRID-trained model.

On the other hand, the relatively poorer performance of the IllustrisTNG-trained model is likely attributed to the fact that IllustrisTNG has suppressed the low-mass galaxy population compared to SIMBA and ASTRID \citep[see for example][]{Ni2023arXiv230402096N,deSanti2023ApJ...952...69D}, resulting in the most sparse stellar maps and making the training more challenging.


\subsection{Recovering the large scale structure in IllustrisTNG300} 
\label{subsection:increase_volume}

\begin{figure*}
    \centering
    \includegraphics[width=1.0\textwidth]{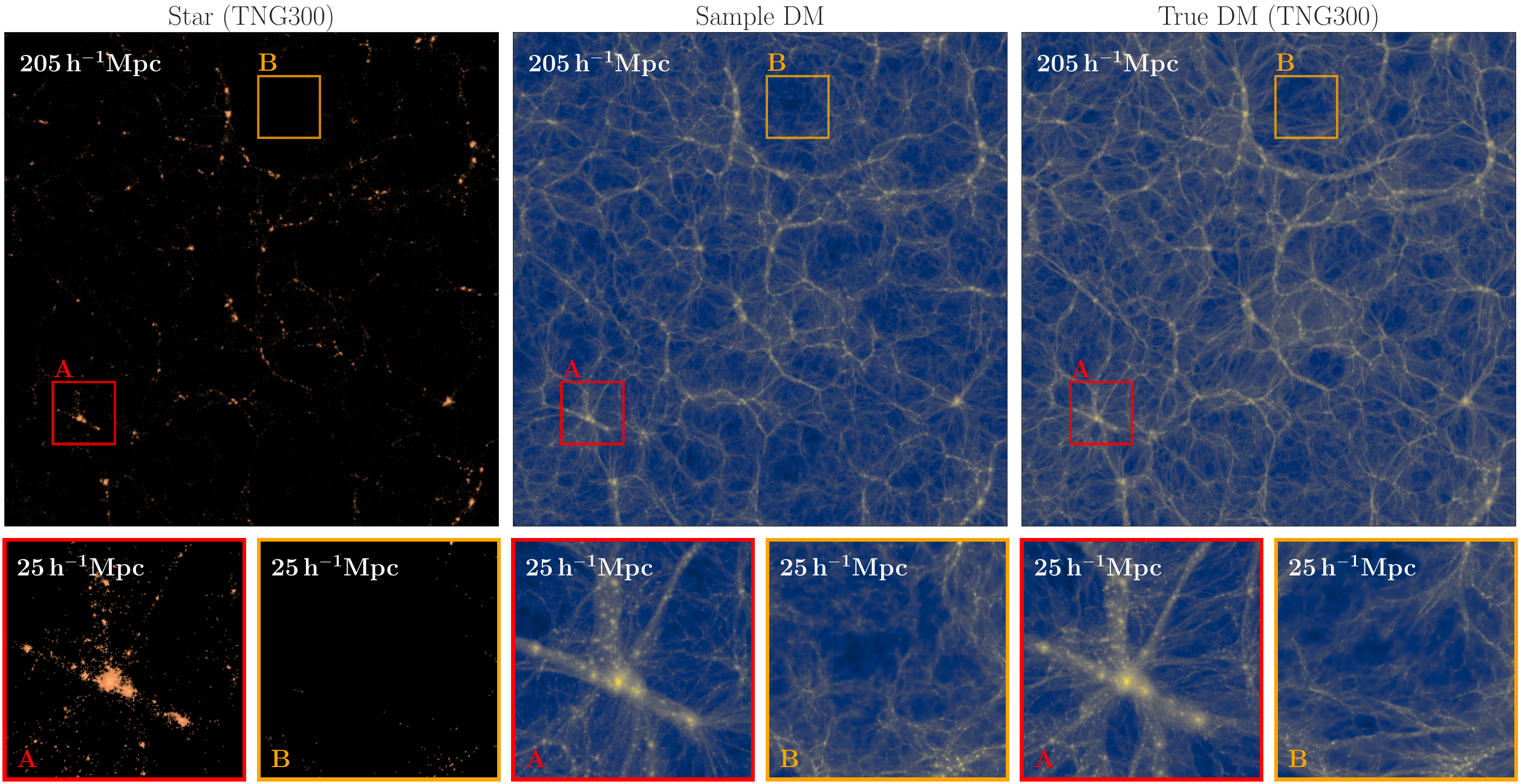}
    ~
    \includegraphics[width=1.0\textwidth]{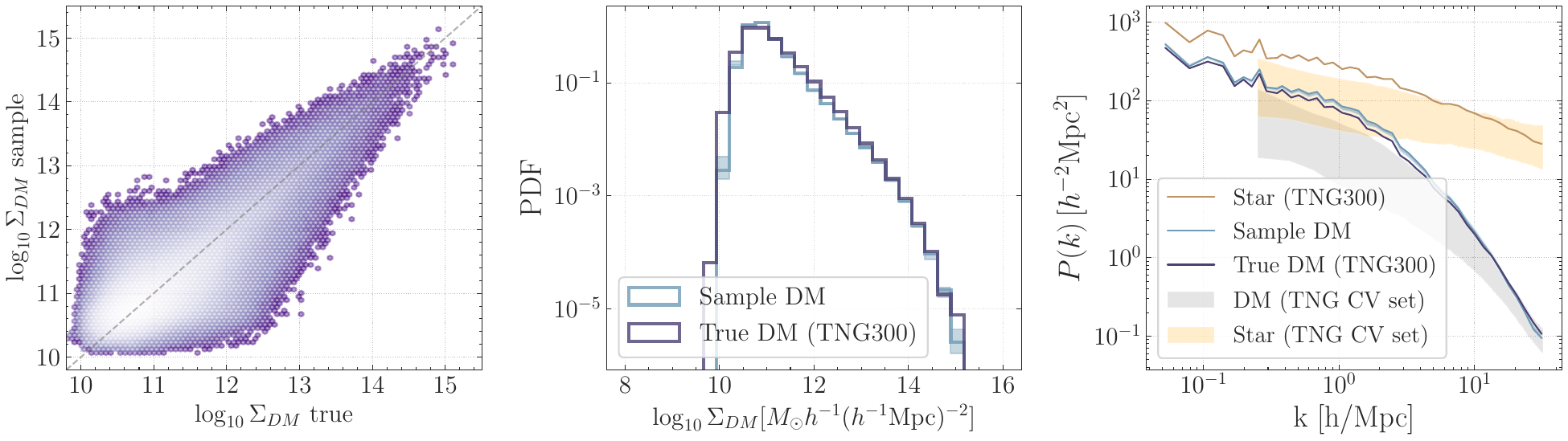}
    \caption{
    We train the diffusion model on the LH set of ASTRID suite and apply it to a much larger volume stellar field from the IllustrisTNG300 simulation. \textbf{Top row}: Full 205 Mpc/h stellar density field, the generated dark matter sample (from our model trained in 25 Mpc/h CAMELS maps), and the corresponding true dark matter density field from IllustrisTNG300. The red and orange boxes highlight 25 Mpc/h sub-regions of a cluster-like region (A) and a void-like region (B). 
    \textbf{Bottom row}: Statistical validations for the 205 Mpc/h fields, illustrating the pixel-level $\Sigma_{\rm DM}$ for both the true and generated dark matter samples on the left, the 1D histogram of the dark matter sample and true $\Sigma_{\rm DM}$ in the middle, and the 2D power spectra of the dark matter and stellar density fields on the right. For comparison, in the rightmost panel we also show the star and dark matter power spectra from the CAMELS IllustrisTNG CV set, which has a lower amplitude due to missing large scale modes.
    }
    \label{fig:debias_tng300}
\end{figure*}

A downside to training our models on the CAMELS simulations is their limited volume, which in particular will be largely affected by super sample covariance \citep{Li_2014}. However, due to the convolutional nature of our model, we can make predictions over much larger volumes. 
In this section, we assess the model's capability to generalize to larger volumes than those in the training set by estimating the dark matter density in the IllustrisTNG300 simulation \citep{nelson2021illustristng,Springel_2017, Pillepich_2017} that has a volume over $500$ times larger than that of the CAMELS simulations used to train the model. 

We preprocess the stellar and DM density fields by maintaining the same spatial resolution as the CAMELS Multifield Dataset, resulting in projected surface density fields of stars and DM with a size of 2100$^2$.
In the upper panel of Figure \ref{fig:debias_tng300}, we show the stellar field over a 2D slab of IllustrisTNG300, together with a sample of the dark matter density field of the diffusion model trained on the small volume CAMELS-ASTRID suite and the true underlying dark matter density field. 
The model can surprisingly produce the correct large scale filaments and voids. Note that we are both extrapolating the trained model to larger volumes and over simulation suites (the model is trained on ASTRID and tested on IllustrisTNG).

In the intermediate panel, we show two highlighted CAMELS-sized regions: a cluster-like region (A) and a void-like region (B). Interestingly, although the void-like region is as large as a CAMELS box, the model still produces a realistic dark matter density field.

Finally, the bottom panel shows the predicted and true dark matter density in each pixel, the density PDF and the power spectra of the samples and the truth, from left to right. 

On the left, we show that although the amount of scatter in the predicted vs. true relation is considerable, we do not see any bias appearing at high values. This is surprising, given that the small volume of the CAMELS simulation limits the appearance of such high density peaks in the training set. 

Moreover, on the bottom right corner, the power spectra of the DM samples agree very well with that of the true IllustrisTNG300 DM density field across all scales, even recovering large scales that extend beyond the fundamental $k$-modes corresponding to the small volume training set.
For comparison, we also show the star and dark matter power spectra from the IllustrisTNG CV set, which exhibit a lower amplitude due to the limited large scale modes in the small-box simulations.

The good agreement of the large volume power spectra demonstrates that the diffusion model (trained on the ASTRID suite) can generalize to (1) different galaxy formation models and (2) scales larger than those contained in the training set.



\section{Summary and Discussion} \label{sec:discussion}

In this paper, we have presented a diffusion generative model that can sample the posterior distribution of DM density fields conditioned on the stellar density field. 
We demonstrate through a diverse set of validation metrics that the generated DM fields are in good statistical agreement with the true DM fields from the simulations.

Moreover, when trained on the LH set of the CAMELS simulation suites, the diffusion model is able to marginalize over the cosmological and astrophysical uncertainties within a given galaxy formation simulation.
Interestingly, the diffusion model exhibits generalization capabilities and can accurately recover dark matter density fields from simulations with alternative galaxy formation models.

Compared to the previous work of \citet{hong2021revealing}, which applies a deterministic convolutional neural network model to learn the mapping between galaxies and dark matter fields based on the IllustrisTNG simulation, our approach is probabilistic in nature, and therefore can capture the inherent uncertainty of the galaxy-DM mapping due to sparsity in the stellar maps, that leads to degeneracies in consistent dark matter distributions and theoretical uncertainties. This in particular allows us to recover posterior samples with consistent small-scale clustering, whereas deterministic models recover a blurry image smoothed on small scales by training the model to reproduce only the posterior mean.

Notably, the mapping between galaxies and dark matter fields can also arise from uncertainties in the cosmological parameters, and more importantly in sub-grid physical models used in galaxy formation simulations. 
Therefore, we train the diffusion model based on the LH set of CAMELS simulations; this dataset features a wide variation in cosmological and astrophysical parameters, allowing the diffusion model to learn how to effectively marginalize over these uncertainties.

We used the 2D projected stellar density field and DM density field from the CAMELS Multifield Dataset as proxies for the galaxy fields and the underlying cosmic web. 
This approximation is, however, rather simplistic and only serves as a proof-of-concept training set to assess the performance of the diffusion model.
To apply the diffusion model to observations of galaxy surveys, future efforts need to focus on making ensembles of simulations with realistic mock synthetic observations of galaxies from simulations, as done in \cite{2023PNAS..12018810H}. 

In the future, we plan to develop a training set targeted at reconstructing the dark matter cosmic web from 3D galaxy point cloud observations of the Dark Energy Spectroscopic Instrument (DESI) survey. In particular, 3D galaxy point clouds are sparser than the stellar mass maps used in this work, and processing them could potentially require a prohibitive amount of GPU  memory.
We plan to assess the feasibility of either doing diffusion in a compressed latent space \citep{rombach2022highresolution}, or developing a diffusion model that can be conditioned on the sparse galaxy 3D point cloud, by using either Graph Neural Networks or transformers as conditioning models \citep{cuestalazaro2023point}. 

Finally, we will include observational effects, such as selection biases, redshift-space distortions, and fiber collisions, in order to train a generative model that can be effectively applied to real galaxy surveys and unravel the cosmic web of our Universe.




\bibliography{references}{}
\bibliographystyle{aasjournal}

\section*{Acknowledgments}
We thank Daniel Eisenstein and Lars Hernquist for useful discussions. 
YN acknowledges support from the ITC postdoctoral fellowship.
This work is supported by the National Science Foundation under Cooperative Agreement PHY-2019786 (The NSF AI Institute for Artificial Intelligence and Fundamental Interactions, \url{http://iaifi.org/}). 
This material is based upon work supported by the U.S. Department of Energy, Office of Science, Office of High Energy Physics of U.S. Department of Energy under grant Contract Number DE-SC0012567. The CAMELS project is supported by the Simons Foundation and NSF grant AST 2108078.
\appendix
\section{Impact of the different galaxy formation models in the stellar density fields}
\label{app:bias}

In this section, we highlight the differences in the bias between stellar density fields and dark matter fields found in the different galaxy models. 

\begin{figure}[!htbp]
    \centering
    \includegraphics[width=0.45\columnwidth]{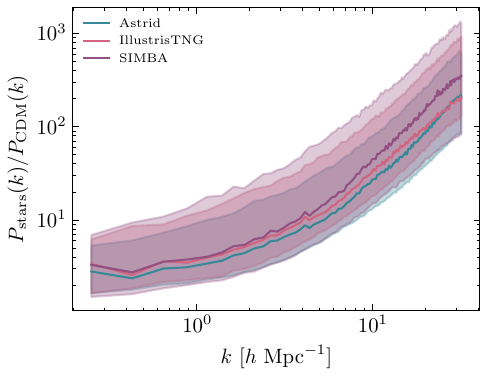}
    \caption{Ratios of star to DM fields power spectra from each dataset, where the shaded region represents the range of ratios for the LH set and the solid line represents the ratio for the CV set. Note that we show the 10-90th percentiles of the LH samples. In general, the different models show a consistent behavior as a function of scale.}
    \label{fig:LH-CV}
\end{figure}

In Figure~\ref{fig:LH-CV}, we show the ratio of star to dark matter power spectra of the different simulation suites. The solid lines show the CV set ratios of each simulation, while the shaded regions represent the 10th - 90th percentiles of the LH samples. In general, the different models show a consistent behavior as a function of scale.

\begin{figure}[!htbp]
    \centering
    \includegraphics[width=\textwidth]{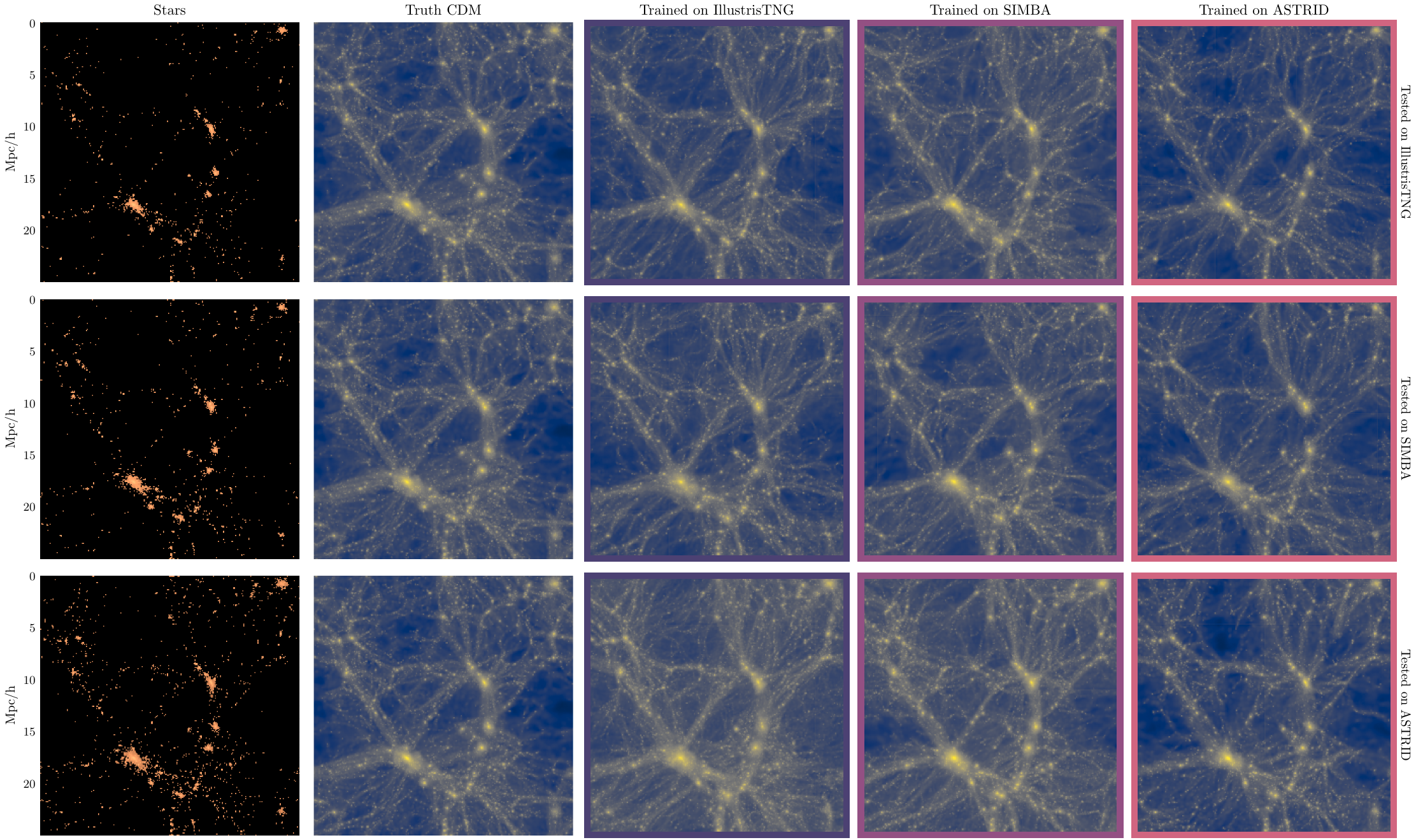}
    \caption{\textbf{First column}: Input stellar fields for each of the simulations suites,  where each row shows the same-index images from the CAMELS CV sets (IllustrisTNG, SIMBA, ASTRID), whose parameter values are the same. \textbf{Second column}: corresponding DM field. \textbf{Third to fifth columns}: DM fields sampled from each model trained on IllustrisTNG, SIMBA, and ASTRID suites respectively. The generated DM fields all correspond well to the true images visually.}
    \label{fig:imshows}
\end{figure}

Moreover, Figure~\ref{fig:imshows} depicts the different stellar fields for the same initial conditions and parameter values in the different simulation suites. It shows that although the dark matter density fields are practically the same in the different simulation suites, the stellar mass maps can look qualitatively different. We also show one sample from the diffusion model trained and tested in all possible combinations.

\section{Ablation Study}
\label{app:ablation}
Deep learning models consist of blocks and parameters that can be adjusted, and often these tweaks present a wide range of predictions. We therefore perform an ablation test to study the behavior of the diffusion model, and examine how the removal or addition of components affects model performance. Our initial model consists of 4 convolution (ResNet) blocks, 48 dimensions of time embedding, and a learned linear schedule. From this model, we implement four other changes varied one at a time: adding an attention block to the bottom of the U-Net architecture, fixing the linear schedule, reducing the number of convolution blocks by half, and reducing the dimensions of time embedding by half. 

Attention is a mechanism introduced by \cite{vaswani2023attention} and is commonly used in deep learning models, namely in natural language processing. In the context of images, attention lets the model highlight only the relevant features of the image. We conduct an ablation with the addition of an attention block to the bottom of the U-Net architecture.

The convolution block is the fundamental block of the denoising model architecture, and four is the typical number used in U-Net, as it is large enough to increase the number of feature channels sufficiently, but small enough to keep the computational resources low. In this ablation test, we halve the number of blocks to observe how the reduced depth of feature channels may affect the neural network's ability to learn the features.

In the forward diffusion process, the model adds noise to the input image in $T=250$ steps, according to a variance schedule. In our original model, we employ a schedule that changes its $\gamma$ parameters in its function as it learns to predict better outputs. In this ablation, we test its performance when we fix the schedule instead.

Time embeddings are how the neural network shares its parameters across time, as each denoising process occurs at each timestep. In other words, $t$ is encoded via time embeddings for the network to know what the current level of the noisy image it is processing at a time. In our original model, we choose 48 as our time embedding dimensionality, and in this ablation we perform one with 24.

We train each model using the IllustrisTNG simulation suite. We train across 300000 steps, or equivalently 300 epochs with 1000 training steps at each epoch, and store checkpoints at every 3000 steps. We choose the metric for this test to be the average cross-correlations, as these are the most fine-grained summary statistics as discussed in Section \ref{subsection:CV-set}.

\begin{figure}[h]
    \centering
    \includegraphics[width=0.5\columnwidth]{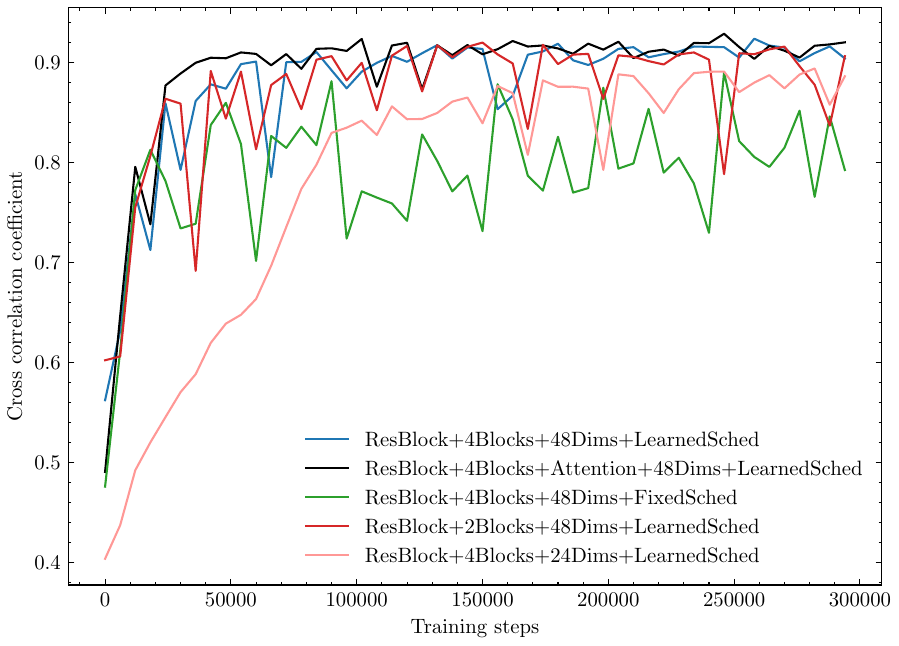}
    \caption{Average cross correlations between true DM and 100 sampled DM fields at every 3000th checkpoint given the same conditioning sample and different diffusion model settings.}
    \label{fig:cross-corr}
\end{figure}

Figure~\ref{fig:cross-corr} shows the performance of each model across training steps. The model with an attention block added consistently yields the highest coefficients, while the one without follows very closely below. Halving the number of convolution blocks shows more fluctuations in their performance, which indicates that the original depth of the architecture is more optimal for it to be stable as the model learns. When using a fixed linear schedule, the model performs well at first, but since it does not learn the best $\gamma$ values, its performance does not improve across training. The model with a half dimension of time embeddings significantly underperforms than the rest, though it catches up by the end of training. 

We conclude from this study that the diffusion model with attention is very efficient in generating accurate DM fields. Since the addition of this block did not need additional computational resources from what we had allocated for all other models, we choose this model as our best, and implement this architecture for further analysis with all simulation suites.

\section{Detailed analysis of  posterior uncertainties}
\label{app:posterior_uncertainties}

\begin{figure}[!htbp]
    \centering
    \includegraphics[width=0.6\columnwidth]{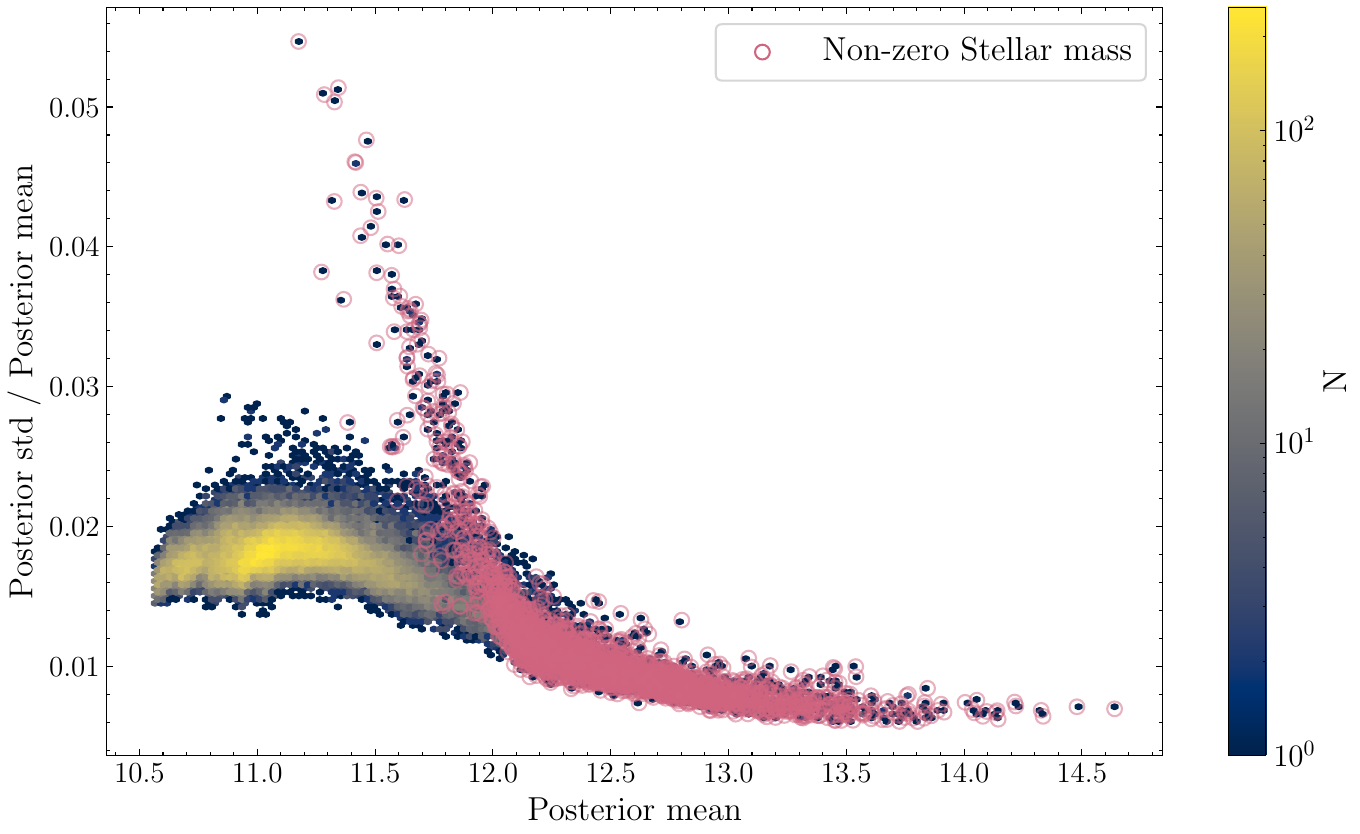}
    \caption{Joint probability density histogram of the posterior ratio of standard deviation to mean, and the posterior mean of the sampled CDM field. All pixels corresponding to locations with non-zero conditioning stellar mass are circled in pink.}
    \label{fig:posterior-stats}
\end{figure}

In Figure~\ref{fig:posterior-stats}, we compare the ratio of posterior standard deviation to posterior mean, as a function of the posterior mean, for one input stellar mass sample from the ASTRID CV set. 

In particular, Figure~\ref{fig:posterior-stats} shows the difference in behaviour for pixels where the stellar mass is non-zero versus pixels with a zero stellar mass. When the stellar mass is low but non-zero, the posterior samples show a higher ratio of variance to mean. This would correspond to small dark matter halos and filaments. On the other hand, when the stellar mass is high, the ratio of posterior variance to mean is the lowest.

\section{Out-of-distribution performance of the parameter inference network}

\begin{figure}[!htbp]
    \centering
    \includegraphics[width=0.32\columnwidth]{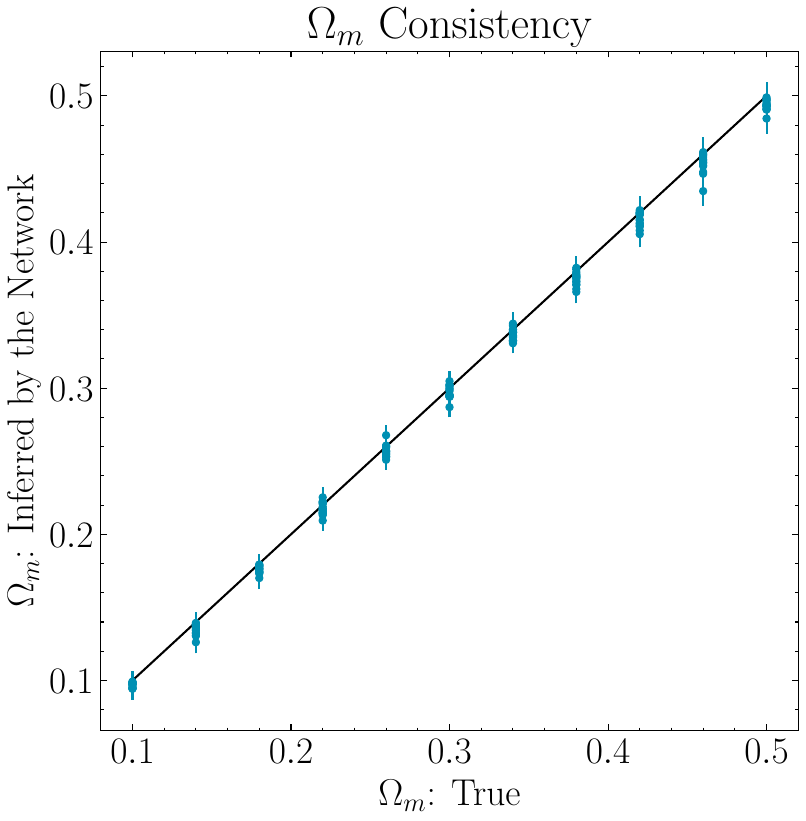}
    \includegraphics[width=0.32\columnwidth]{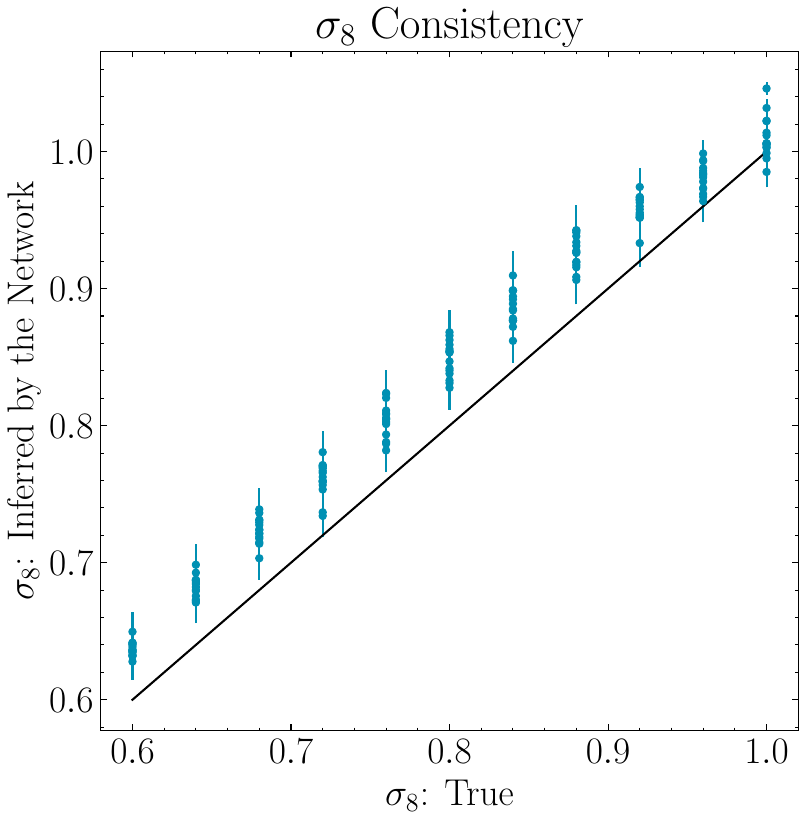}
    \caption{Performance of the parameter inference networks from \cite{2021arXiv210909747V} on the true 1P DM fields for ASTRID. On the x-axis, we show the true $\Omega_m$ and $\sigma_8$ values for each simulation. The y-axis shows the mean and variance of the posterior predictions for the cosmological parameters given a dark matter density field from the ASTRID LH simulations.}
    \label{fig:infet-true}
\end{figure}

In Figure~\ref{fig:infet-true}, we show the predictions of the parameter inference network, trained on a subset of the Illustris-TNG LH set to infer $\Omega_m$ and $\sigma_8$ given a dark matter density field, when tested on the dark matter density fields from the ASTRID LH set. In particular, Figure~\ref{fig:infet-true} demonstrates that the inference network produces biased values of $\sigma_8$.

\end{document}